\documentstyle[seceq,epsf,preprint]{ptptex}

\title{
Getting Ready for GEO600 Data
}

\author{
Bernard F. {\sc Schutz}\footnote{E-mail address: schutz@aei-potsdam.mpg.de}
}

\inst{
Albert Einstein Institute, 14476 Golm, Germany, and \\
Department of Physics and Astronomy, Cardiff University, Wales
}

\recdate{
}

\abst{
Data of good quality is expected from a number of gravitational wave 
detectors within the next two years.  One of these, GEO600, has 
special capabilities, such as narrow-band operation.  I describe here 
the preparations that are currently being made for the analysis of 
GEO600 data.}

\preprintnumber{AEI-1999-26}

\begin{document}

\maketitle

\section{Introduction}
After 4 decades of development, gravitational wave detectors are finally 
reaching the level of sensitivity where the first detections could occur. 
Within one year, the TAMA\cite{TAMA} detector could detect short bursts 
of radiation with an amplitude smaller than $10^{-20}$, and by 2001 both 
LIGO\cite{LIGO} and GEO600\cite{GEO} should be sensitive to bursts of 
amplitude around $10^{-21}$. Within a further year, VIRGO\cite{VIRGO} could 
come on-line with an even better sensitvity.

Detections at this level are by no means certain, but a strong supernova 
in a nearby galaxy could be seen. In addition, these 
detectors can be used to look for other sources with smaller amplitudes. 
Since their sensitivity increases roughly as the number of cycles of 
radiation they observe, a long-lived source like an irregular spinning 
neutron star could be detected with an amplitude as small as $10^{-26}$ in 
one year of observing. 

The possibility that these detectors can be used to search for different kinds 
of signals makes them very versatile, but it also complicates the analysis 
of the data. Each detector produces a single data stream that may contain 
many kinds of signals.  Detectors don't point, but rather sweep their broad 
quadrupolar beam pattern across the sky as the Earth moves.  So possible 
sources could be anywhere on the sky.  Data analysis algorithms need to 
be able to find signals from any location.

Since the first detectors will barely cross the threshold of detectability, 
the early detections will be weak, with small signal-to-noise ratios. There is 
therefore a premium on applying the best signal analysis methods to this 
detection problem.  The best linear method, for example, is matched 
filtering, in which the computer looks for a correlation in the noisy 
data with a template, which is an expected waveform.  For some sources, 
such as coalescing compact-object binary systems\cite{cbs} 
and spinning neutron stars\cite{Brady1997}, 
we believe we have good theoretical templates.  However, in all cases 
the waveform templates depend on parameters, such as the masses of 
the binary stars or the location on the sky of the neutron star, that may not 
be known ahead of time.  Using these templates becomes a compute-intensive 
job, and the algorithms must be implemented as efficiently as possible.

For other sources, such as the last phase of the merger of 
two black holes\cite{Seidel} 
or the burst of radiation from the gravitational collapse event that leads 
to a supernova,\cite{Eanna}  our understanding of 
the astrophysics does not at present 
permit us to construct accurate templates.  Other methods that are more 
robust than matched filtering must be used in order to recognize signals 
of an unexpected shape.  Such methods must be chosen with care, since 
they are not optimal and some may perform better than others on specific 
sources.

In this paper I will review the progress being made within the GEO600 
project to design a data analysis system that is capable of 
performing sensitive searches for all of these 
sources within the project's limited computing budget.  The 
data analysis programs are being developed jointly with scientists 
in the LIGO Science Collaboration (LSC) and with 
VIRGO scientists, so much of what I say 
will also apply to those projects.  But GEO600 has, unlike the other 
first-generation detectors, the ability to do signal recycling,\cite{sigrecyc}
which is a form of narrow-band observing.  I will discuss the impact 
of signal recycling on our plans for data analysis.

Since expected signals are weak, there has been significant interaction 
between designers of the experimental hardware and data analysis 
specialists in GEO600.  I will give examples of how some experimental 
features were modified or improved, and others specified in detail, 
after studies of data analysis showed the benefits of taking these steps.

The plan of the paper is as follows. After a review of the current status of 
the GEO600 detector, I will outline the plans for data analysis, both within 
GEO and in collaboration with LIGO.  I will then address the detection 
algorithms needed for specific sources: chirps, known pulsars, searches 
for unknown neutron stars over wide areas, the X-ray binary Sco X-1, 
and unexpected signals.  The first source detected might fall in 
any of these categories, and it might come at any time from the moment 
detectors begin operating.

\section{Status of GEO600}
GEO600 is a collaboration among three institutions in Germany --- the Max Planck Insitute for 
Quantum Optics, the Max Planck Institute for Gravitational Physics (Albert Einstein Institute --- AEI), and 
the University of Hannover --- and two in the UK --- the University of Glasgow and Cardiff University.
The detector, with 600~m arms, is under construction near a village called Ruthe, south of Hanover. 
Its vacuum system is fully constructed and tested, the first mode cleaner cavity is locked and 
working, and the data acquisition system is installed.  Currently work is going on to lock one 
arm as a single optical cavity, and this is planned for the end of 1999.  By mid-2000 there should 
be interferometry with the test optics, and full interferometry with the final optics the following year.

\begin{figure}
\label{fig:senswide}
\epsfxsize = 9 cm   
\epsfclipon
\centerline{\epsfbox{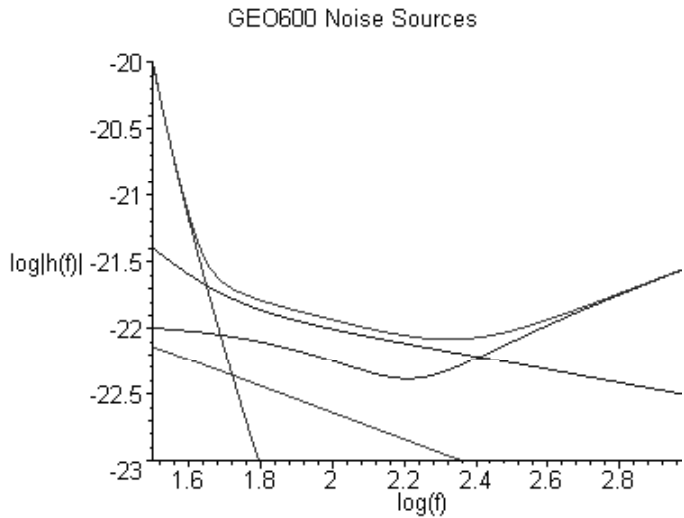}}
\caption{GEO600 expected noise curve in broadband mode.}
\end{figure}

\begin{figure}
\label{fig:sensnarrow}
\epsfxsize = 9 cm   
\epsfclipon
\centerline{\epsfbox{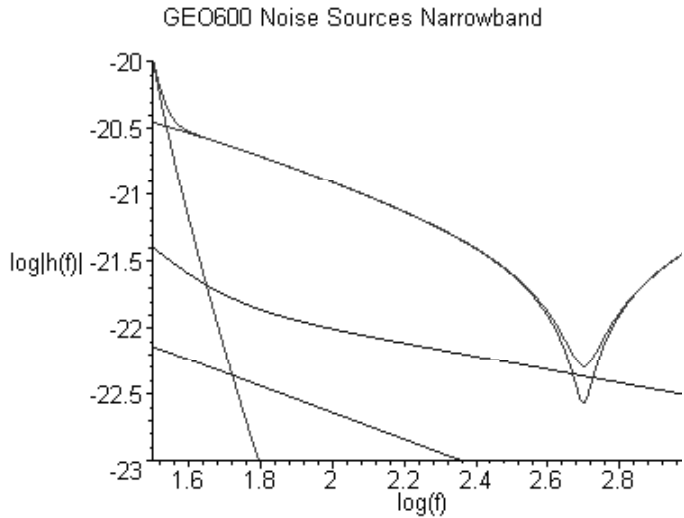}}
\caption{GEO600 expected noise curve in a particular narrowband configuration.}
\end{figure}

Figure~\ref{fig:senswide} shows the expected performance of GEO600 in its broadband configuration 
and Figure~\ref{fig:sensnarrow} in a possible narrow-band configuration.  The total noise is shown as a quadrature sum of 
the expected noise from different sources.  Notice that the photon shot noise even in the broadband 
case is tuned by signal recycling.  This allows the shot noise to be adapted to the best form 
to fit with other noise sources, and it also makes the interferometry cleaner.  The narrow-band 
curve on the right is shown for a specific frequency near 500~Hz.  The bandwidth is selected to 
allow the best sensitivity to be limited only by the thermal noise.  This mode is useful in looking 
for signals of known frequency, or in performing searches for narrow-band signals at high 
frequency.


The GEO600 project has its own data-analysis team inside the project, 
distributed among Cardiff, the AEI, and Glasgow. This is in contrast to LIGO, which 
provides support and coordination for data analysis within the project but relies  
on the LSC to provide the algorithms and much of the software.  The 
present author is the principal investigator with responsibility for 
data analysis and acquisition within GEO.  B.~Sathyaprakash at 
Cardiff has overall responsibility for data analysis coordination, 
and he is assisted by research scientist I.~Taylor, 
postdoctoral scientists B.~Balasubramanian, 
and D.~Churches, and programmer M.~Lewis.  
 
At the AEI the current GEO support group 
includes staff members 
C.~Cutler, M.-A.~Papa, and A.~Vecchio, plus postdoctoral 
scientists E.~Chassande, B.~Owen, A.~Sintes, C.~Ungarelli and 
P.~Williams.  The staff at the AEI work on a mixture of roughly 50\%
preparation for data analysis and 50\% theoretical work on 
gravitational wave sources, including template prediction. 
At Glasgow a new group consisting of staff member C.~Davies 
and postdoctoral fellow P.~Boyle is expected to grow by 
at least one further member soon.  In addition, the GEO team 
has strong collaborations with individuals outside the 
project, including S.~Frasca (Rome), A.~Krolak (Warsaw), and  
S.~Dhurandar (Pune).

\section{ Approach to data analysis}
\subsection{Data storage and distribution}
Data from dozens of control, monitoring, and signal points 
will be acquired on the site and distributed around the site or to more distant 
points in real time, as required by the experimentalists.  The experiment 
site is connected to Hanover University by a high-speed radio link, 
so no data is stored at the site (apart from a disk buffer capable of 
storing 3 days of the archive data set). Twin redundant workstations 
control data distribution from Ruthe to Hanover and beyond.

In Hanover a cluster computer with a total processing capacity of 
about 5~Gflops will clean and calibrate the data, and 
perform initial searches for burst sources and coalescing 
binaries.  Wide-area searches for signals from unknown 
neutron stars will be performed on a cluster computer at 
the AEI, which should have a speed approaching 20~Gflops, 
with 40~GB main memory.  A similar cluster is planned in 
Cardiff, which will perform similar searches for different 
ranges of parameters.  We are currently testing 
CPU chips to determine what kind of machine would be 
best suited for these clusters.

A subset of the total 
available data will be archived to tape at $0.5\;\rm MB\;s^{-1}$.  
This may be done in Hanover, but we are currently examining 
the possibility of sending over the internet to Berlin and 
writing it to tape there.  This archive, some 15 terabytes per year, 
will be stored in two places: in Germany, by the Albert Einstein Institute, and 
in Britain by Cardiff University.  It will be available to all members of 
the GEO project, all close collaborators, and (under an agreement 
described below) to members of LIGO and most of the LSC, 
but the size of the data store suggests that it will not often 
be accessed, nor will it be distributed to other sites.

By contrast, most of the real work will be done using a reduced set of data, 
containing the cleaned and calibrated 
signal data and a summary of the monitoring and environmental data. 
This will be the foundation of most of the subsequent data analysis 
within GEO, and it will 
be distributed to all the GEO partners, collaborators, and LIGO. The reduced 
data set will 
probably amount to a much more manageable size of 0.5 terabytes per 
year.

\subsection{Software}
Within GEO we have been designing two kinds of data analysis software.
The first is quick-look software for commissioning, which will help scientists 
track down trouble spots and diagnose them as they try to get the detector 
working.  The second is our main data analysis system.

These two kinds of analysis have different requirements.  For the quick-look 
system we have designed a new computational environment called Triana\cite{triana}.
A preliminary version of this software has been running for almost a year at the 
Caltech 40~m prototype interferometer in the LIGO project, and is now running 
at the Ruthe site for GEO.  This software, written in Java, has a simple graphical user 
interface, in which a user assembles a data analysis program from a set 
of pre-programmed components (FFTs, correlators, etc) by dragging the 
components from a toolbox into a workspace and then wiring them together 
to direct the data through them.  The rapid re-configurability of the system, its 
intelligent built-in data types, its ability to work over networks, its 
ability to take several data streams from the acquisition system at the 
same time, and its suite of powerful applications take it well beyond 
the usual functions of  
oscilloscopes and spectrum analyzers.  Using Triana and the 
computer-based control system manager, scientists will be 
able to sit in Glasgow or Hanover and troubleshoot technical 
problems.  Although Cardiff University will be selling Triana 
commercially, it will be free to the gravitational wave community 
from the Triana website.\cite{website}  The first public release 
is expected in late 1999.

For the full data analysis programs, Triana is not a full solution.  Programs must 
be written in a compiled language (C or FORTRAN) for maximum speed, and 
algorithms need to be optimized for efficiency; ease of reconfigurability is not 
a requirement.  Nevertheless, Triana is planned to be used by GEO for the 
top level, supervisory program.  This is because this program does very little 
computation, so Java is no disadvantage. Triana's efficient networking abilities 
allow the program to be controlled from, say, Cardiff while it runs at Hanover, 
despite the relatively slow internet bandwidth between the two locations. And 
finally, Triana's built-in data types, which were designed for looking quickly at 
the data from the detector, are perfectly suited for passing information between 
steps in a larger analysis algorithm.   So the present plan is to use Triana to
control and manage the communication between the C/FORTRAN functions 
or subroutines that will perform the data analysis.

\subsection{ Cooperation with LIGO}

GEO and LIGO have signed an MOU that provides for full access from each 
project to the other's data.  This will last as long as the LIGO~I project 
takes data.  Rules are currently being discussed for managing the access 
and exchange of the data.  The expectation is that any members of the projects 
or their collaborators (eg LSC members who contribute to LIGO~I) should 
be able to get access to the data of both projects for analysis.  Publication 
on the basis of such analysis would have to be approved by both 
project managements, and important papers would have authorship 
consisting of the entire collaborations.

Because of the size of the data sets, full access to non-signal data 
streams, such as environmental monitors and so on, will only 
be practical at the archive centers. But the smaller signal data 
streams will be distributed and shared between the projects.

This agreement is not exclusive. Both projects see this as a model 
for an enlarged agreement that will hopefully involve TAMA and VIRGO as
well. The goal of gravitational wave astronomy is best served by 
data exchange and joint observing programs.

\section{ Specific sources: sensitivity, algorithms, priority}

\subsection{ Chirps }
The first interferometers will not have good sensitivity for detecting coalescences of 
neutron-star binaries, unless there are orders of magnitude more such systems 
in nearby galaxies than in our own. Figure~ref{fig:cbs} shows the sensitivity of 
GEO600 to a neutron-star chirp in the Virgo cluster.  This is not really detectable, 
yet on current statistics only one such  event would be expected every 1000 years.
We will therefore have to wait for detectors in the class of  
LIGO~II before we see these.  
\begin{figure}
            \epsfxsize = 9 cm   
		\epsfclipon
            \centerline{\epsfbox{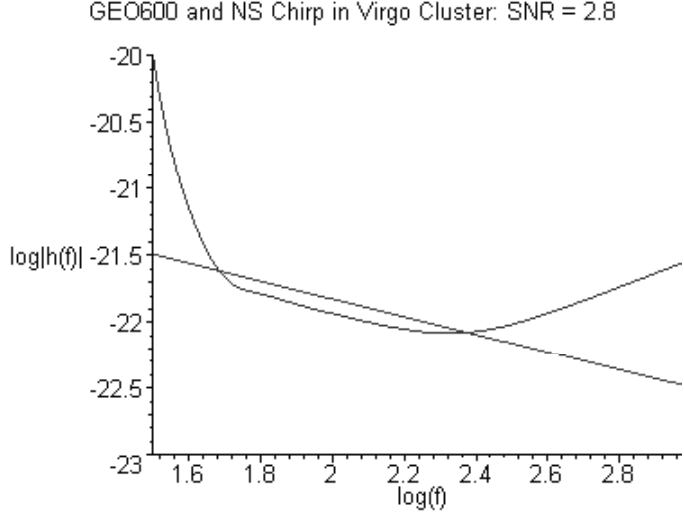}}
        \caption{Sensitivity of GEO600 vs the effective signal from a 
binary system consisting of two $1.4\;M_\odot$ neutron stars in the Virgo 
cluster. The noise is plotted as $S(h)^{1/2}$.  The signal is 
plotted as $f^{1/2}|\tilde{h}(f)|$. In this form, the difference between the 
curves in this logarithmic plot is a visual indication of the contribution 
to the SNR from any frequency. In this case, the signal is barely above the 
noise, leading to a SNR of 2.8.  This is not enough to identify a detection 
without other coincident observations.}
        \label{fig:cbs}
        \end{figure}

\begin{figure}
\label{fig:bhcbs}
\epsfxsize = 9 cm   
\epsfclipon
            \centerline{\epsfbox{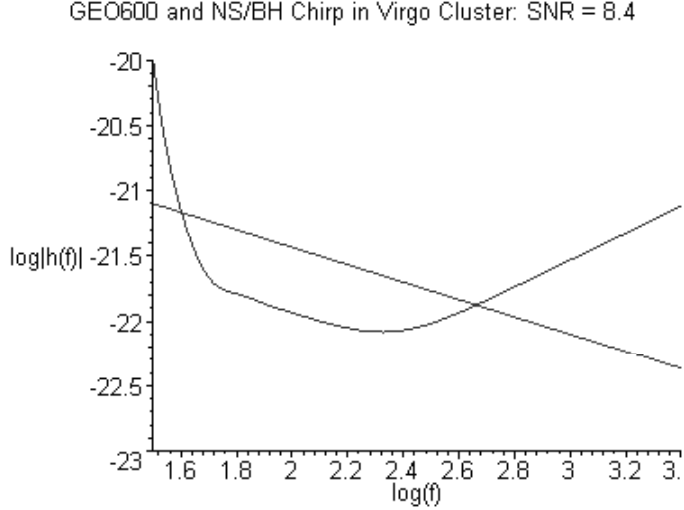}}
\caption{Signal and noise for GEO600 and a binary coalescence consisting 
of a $1.4\;M_\odot$ neutron star and a $15\;M_\odot$ black hole in the Virgo cluster.}
\end{figure}

But other binaries may be seen earlier.  We don't 
know the rate of black-hole/black-hole binaries. Being stronger, if they are as 
abundant as neutron-star binaries, then there may be one or two near enough to 
the Earth to be detected by the first detectors.  Figure~\ref{fig:bhcbs} shows, 
for example, that binary stellar black holes could be visible to GEO600 out to 100~Mpc with 
SNR of 5.  Given the small numbers on which rate estimates are based, this 
offers a real possibility of detection.  But if the black holes are even more massive, then things 
get less certain.  The observable frequency region may include only  the merger radiation itself, and 
we don't have good models of this at present. Detection will rely on more robust methods.\cite{Eanna}
\begin{figure}
\epsfxsize = 9 cm   
\epsfclipon
            \centerline{\epsfbox{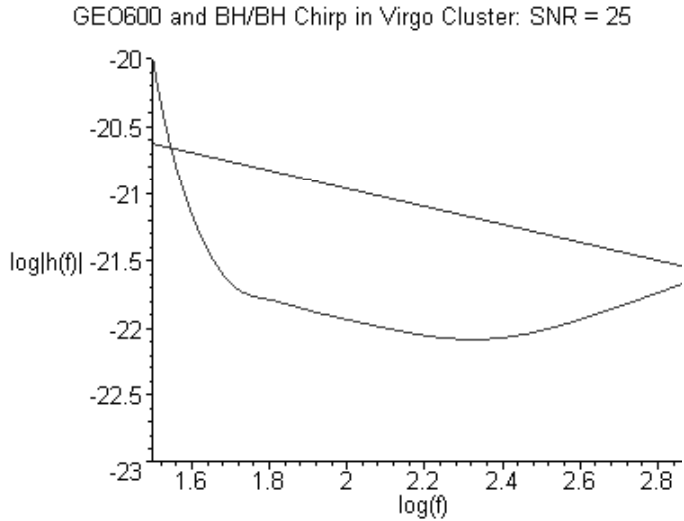}}
\caption{ Signal and noise for GEO600 and a binary coalescence consisting 
of two $15\;M_\odot$ black holes in the Virgo cluster.}
\end{figure}

An interesting idea is the possibility of 
binary systems formed by MACHOs.\cite{MACHO}  If these objects are compact enough to 
generate gravitational waves in the observable region, then these binaries could 
be very abundant.  For this reason, GEO is planning to search for binaries with 
a total mass as low as $0.5\;M_\odot$.

Correlations with unusual events may also help, such as gamma-ray bursts. But 
the rate of bursts suggests that, if they are correlated with neutron-star 
binaries, they are quite far away.  However, if bursts can arise from a number 
of different systems, then some, such as hypernovae, may be nearer.

\subsection{Continuous waves from known objects. }
If a pulsar is known through radio or X-ray observations, then searching for it 
in gravitational wave data will normally not be very challenging.  The position 
of the source will be known, and the frequency of the radiation must be 
closely related to the pulsar frequency.  It will be important to search 
at frequencies of the pulsar frequency ($m=1$ quadrupolar radiation, 
which might come from free precession), twice the pulsar frequency 
($m=2$ quadrupole radiation, which would be generated by mass 
asymmetries built into the spinning star), and 4/3 the pulsar frequency 
(which would come from an unstable $r$-mode.)  Indeed, free precession 
can make these frequencies differ by small amounts from the pulsar 
frequency, so searches will have to be done in small bandwidths about 
these key mulitples of the spin frequency.

But to achieve a reasonable sensitivity we must accumulate signal 
for a year or more.  This places constraints on the operations of 
detectors.

\begin{figure}
\epsfxsize = 14 cm   
\epsfclipon
            \centerline{\epsfbox{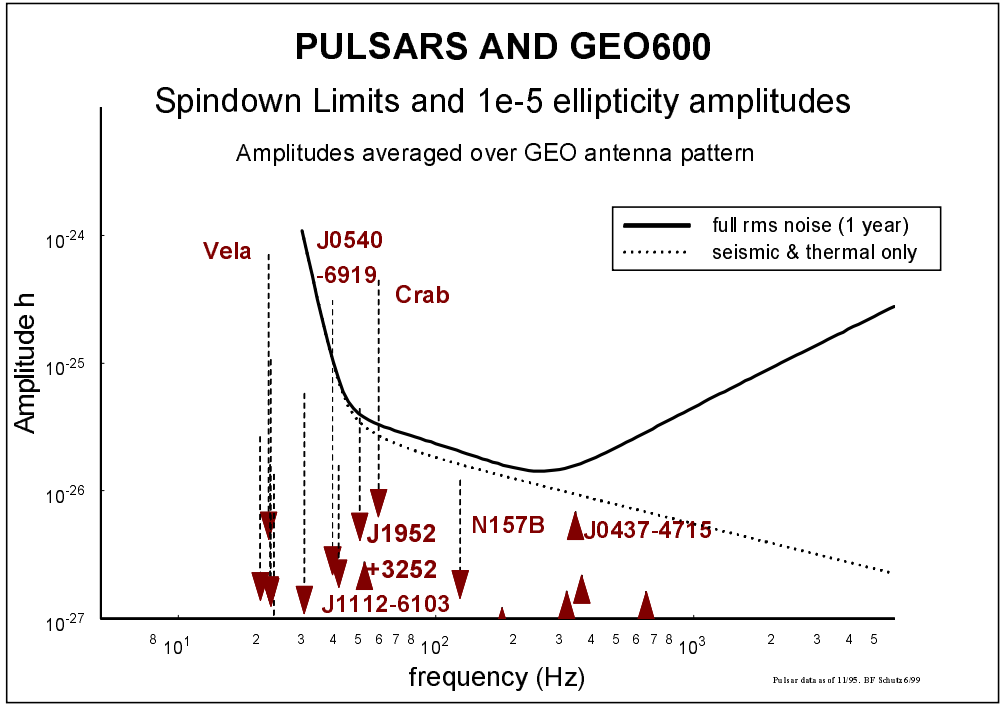}}
        \caption{Noise-curve of the GEO600 detector in broadband 
mode for a one-year observation, compared to signal amplitudes 
from a number of known pulsars. The amplitudes are shown as 
arrows, the top of which is at the spindown limit and the bottom 
at the amplitude expected for an ellipticity of $10^{-5}$.  The 
amplitudes are averaged over for the antenna pattern of GEO 
as it sweeps across the sky.}
        \label{fig:psrs}
        \end{figure}

Known pulsar amplitudes are also constrained if the period 
derivative of the pulsar is known, as it is for most pulsars. The 
loss of kinetic energy determine by the spindow sets a maximum 
amplitude on the expected radiation, provided we know also 
the distance to the source. In most cases this maximum is 
itself below the GEO observing curve, but in a few cases
this limit is interesting.  These are illustrated in Figure~\ref{fig:psrs}.
The most important of these cases is
the Crab pulsar, which GEO could see with SNR more than 
100 in one year, provided it is radiating at its limit.  However, 
if an eccentricity of about $10^{-5}$ for the source of the 
radiating quadrupole moment is assumed, then the radiation 
amplitude falls to a little below the GEO noise curve. Only 
observations will tell whether the Crab is detectable by GEO. 

A complicating issue for the Crab pulsar is the fact that it 
glitches relatively frequently, so that the possibility that 
a glitch will occur during a one-year observation is 
significant. The effect of a glitch is to change the frequency 
of the radiation, but if the radiation depends on an asymmetry 
that is changed by the glitch event, then also the phase of 
the gravitational waveform could undergo a sudden change. 
This means that filtering for the Crab will not be trivial. We 
will have to use continuous radio observations to monitor 
the frequency, and if it glitches we will have to put in 
a parameter in the filter for the phase jump.

The Crab illustrates the interaction within the GEO 
project between experimentalists and data analysts. 
A year ago, the GEO noise curve would have been drawn with 
its low-frequency limit (steeply rising seismic wall) about 
10~Hz higher than is shown here.  After the data 
analysts emphasized to the experimenters the 
potential importance of achieving good sensitivity 
to the Crab, the experimenters re-designed the suspension 
to improve its sensitivity at the crucial frequency of 60~Hz 
by a factor of almost 100.  This new design was so 
successful that it has been adopted as part of the 
design for the LIGO~II upgrade.

Other known objects may be sources of gravitational 
radiation, but we might not have as much information 
from them as from pulsars. For example, unusual 
giant stars may be Thorne-Zytkow objects, which 
are giants with neutron stars in the centers, the 
result of inspiral of a binary neutron star companion.
The neutron star will be accreting strongly and 
may, if the giant rotates, accrete angular momentum 
and spin up.  This could lead to instabilities or 
other gravitational wave radiation mechanisms, 
as proposed for the low-mass X-ray binaries. 
In such systems we will have a location to search 
for, but not a frequency.

Supernova remnants should also be searched. Young 
remnants may contain a neutron star that is a pulsar 
that is not beamed toward us.  Recent observations 
by the Chandra satellite have revealed a previously 
unobserved point source of X-rays at its center, 
which is likely to be a young neutron star.  GEO should 
search all such remnants in a small area around 
their geometric centers.

Globular clusters are promising sources of radiating 
neutron stars.  Many pulsars have been found in 
a few core-collapse globular clusters. Some are so 
young that they must have formed by accretion-induced 
collapse of white dwarfs.  They may be sources of 
gravitational radiation themselves, but GEO should 
search the rest of the field of such clusters for 
other neutron stars.  These might be pulsars beamed 
in other directions.

GEO is developing a database that will include the 
coordinates of all such targets.  The selection of 
systems to look at can be done at any time.

\subsection{ Area searches }
Pulsar surveys do not show us all the neutron stars in 
the Galaxy, because of beaming effects and the limited 
range of such surveys. Gravitational wave observations 
have the potential to be more complete, at least in 
terms of strong gravitational wave emitters.  But to 
become this complete, we shall have to do a wide-area 
survey of the sky.

The youngest star in the galaxy is probably about 10\% of 
the age of the youngest known pulsar, the Crab.  It may be 
nearby, hidden in a star-forming region, or it may be on 
the other side of the Galaxy.   If young neutron stars are
strong radiators, then this star should be the strongest. 
According to recent work on the $r$-mode instability,\cite{rmode} 
it should not be rotating faster than about 150~Hz, 
so its radiation should be at or below 300~Hz, where 
GEO and other first-generation detectors have 
optimum sensitivity.  But its frequency will not be constant 
during the observation.  If it has a spindown timescale of 
only 100 years, then in a one-year observation it will 
spin down by 1\%, or about 3~Hz.  Since we can resolve 
$3\times10^{-8}$~Hz in a one-year observation, this 
spindown represents a drift in frequency of $10^8$ 
frequency resolution elements (frequency bins)!  
The drift would be noticeable in just 
a one-hour observation, too short to accumulate enough 
sensitivity to detect radiation of a reasonable amplitude.

In fact, the situation is worse.  if the second time derivative of the frequency 
acts on the same timescale, then it makes a change of 
frequency of 0.01\%, or about $10^6$ frequency bins.  To 
fit the signal over one year would require including  
the third time derivative ($10^4$ bins) and the
fourth time derivative (100 bins).  Each different frequency bin 
that can be affected by the spindown parameters (derivatives of 
the frequency) 
forces one to use a different value of that parameter that needs 
to be searched for, so just to accommodate the spindown 
one needs roughly $10^{20}$ templates!

In addition, its position will be unknown.  In a 1-year 
observation, the frequency modulation caused by the motion 
of the detector allows one to obtain a similar angular 
resolution on the sky to that which radio astronomers 
obtain with single-dish observations of pulsars, namely 
fractions of an arcsecond.  That means that there are 
some $10^{13}$ locations on the sky that must be 
searched independently in order to demodulate the
signal well enough to reach the theoretical sensitivity 
of such an observation.  For each place on the sky 
one has to look through the full set of spindown 
parameters.

So a brute-force search for such a pulsar will have to cope with 
a parameter space consisting of some $10^{33}$ 
templates, any one of which might reveal this pulsar at 
any frequency. 
But it is easy to see that to perform demodulation and 
a wide-band frequency search for this many templates on a year's worth 
of data sampled at, say, 1~kHz ($3\times10^{10}$ samples) 
is simply impossible.\cite{Brady1997}  

But searches for such pulsars are not hopeless. It is 
clear that if one performs $10^{33}$ independent searches 
with approximately $10^{10}$ independent frequencies in 
each search, 
then one cannot expect to identify a pulsar right down to 
the $1\sigma$ noise level of the detector: one can only 
be confident of an identification if the probability that the 
noise would create a chance signal is less than about $10^{-43}$. 
Given Gaussian noise, this requires a threshold for 
identification of at least SNR = 9.3.  It seems safe to 
assume that such a search will have an effective sensitivity 
no better than 10 times the detector noise level.  So 
a search does not need to examine each frequency bin 
down to the $1\sigma$ level.  It only needs to ensure 
that a signal as strong as $10\sigma$ is not missed.

Hierarchical search strategies can address such a 
requirement much more efficiently, provided they 
are carefully designed.  Two related methods are 
under development, one in the LIGO LSC community,\cite{BradyCreighton}
and the other within GEO.\cite{Papa}  Both begin 
from much shorter initial searches, lasting approximately 
1 day.  Since such a search has frequency bins 365 times 
larger than those of a 1-year search, and the modulation
produced by the Earth's motion and by spindown is much less, 
the number of independent templates that must be searched 
is very much less, about $10^7$ instead of $10^{33}$.  To 
perform a complete search over all these parameters in one 
day (keeping up with the data stream) requires a computer capable of 
executing a few hundred gigaflops.  The methods then 
knit together a succession of these short searches using 
an {\em incoherent} method.  In the LIGO approach, 
this is done by adding power spectra together. In the 
GEO approach, this is done by searching power spectra 
for patterns of peaks, using a technique developed in 
particle physics called the Hough Transform.  

At the end of this incoherent stage, the data are passed 
through a threshold chosen so that a signal that would have 
had a strength of $10\sigma$ would still be seen.  This 
turns out to be at a level of $10\sigma/365^{1/4}=2.2\sigma$ 
at the end of the incoherent stage.  Then a third stage 
is required to follow up these candidates by refining 
parameter space, but only around the parameters that 
pass through the threshold, and then repeating the search 
over this smaller portion of parameter space. Since 
the parameter space is smaller the initial coherent 
stage of the next search can be longer, and so by the 
end of the second incoherent stage the threshold can 
be set to produce a further refinement of the space.  In 
this way one can iterate to eliminate all the parameter 
space in which there are false alarms.

Even this search, just described, demands considerable 
computer power and is not at all optimal.  Practical 
searches will probably have as a goal finding sources 
provided they are stronger than some target level that is  
somewhere between $10\sigma$ and $15\sigma$ over 
an observing period of  only $10^7$~s.  Optimization studies 
by GEO\cite{Papa} have suggested this is possible for some kinds of 
sources with a 20~gflops computer.  Further studies are 
needed when the algorithms are coded in the next few months.

Such algorithms are limited not just by the availability of 
computer cycles but also by considerations of memory size 
and input/output delays.  In GEO we have developed methods 
of approximating Fourier transforms of data over long periods, 
such as one day, by building them up out of Fourier transforms 
of data taken on short periods of time, say an hour, and working 
exclusively within a narrow bandwidth of frequencies.\cite{schutz}  While 
not exact, our tests have shown us that such methods produce excellent approximations, 
and have the advantage that a single processor in a parallel computer can work 
entirely locally in frequency space, without needing to communicate 
with other processors.  Because of these methods, GEO plans to 
achieve its analysis targets using clusters of workstations 
communicating only via fast ethernet links.

Further improvements in these algorithms seem possible. 
At frequencies above 300~Hz, narrow-banding is possible. 
This improves sensitivity in a narrow band of frequencies, 
allowing one to reach the same limit on the signal amplitude 
in a shorter time.  Since the number of parameter sets is 
strongly dependent on the observation time, it is actually 
a computationally more efficient strategy to cover the 
high-frequency region in a number of short narrow-band 
observations than in a single wide-band observation. 

Another improvement will be to run the GEO-designed 
Hough algorithm alongside the LSC-designed power-spectrum-addition 
algorithm.  Both have similar sensitivity and performance, 
but they probably treat the noise in different ways.  By demanding 
that a candidate source should pass both tests simulaneously, 
we should be able to further narrow the volume of parameter 
space in the follow-up stages, improving the sensitivity of the 
overall search.

\subsection{Sco X-1}
One of the surprizes of the last few years in X-ray astronomy has been the
discovery that the spin frequencies of the neutron stars in low-mass X-ray binary 
systems lie in a narrow range around 300~Hz.\cite{lmxb}  Since these 
are accreting neutron stars being spun up to the millisecond pulsar range, it 
would be reasonable to have expected a distribution of spins.  Some mechanism 
must exist to limit the spins of the stars to this particular value.  Magnetic effects 
do not seem to be a good candidate, since the stars have different accretion rates, 
so equilibrium with the magnetic field at the same spin for all stars would require 
the magnetic field to be correlated with the accretion rate.  

A different suggestion has been made by Bildsten.\cite{bildsten}  He suggests that 
gravitational radiation limits the spins, and in particular that accretion leads to an 
asymmetry in the temperature of the star, which in turn leads to an asymmetry in 
the density distribution.  The resulting quadrupole moment would emit gravitational 
radiation, and at some spin this would carry off enough angular momentum to balance
that which is accreted. Detailed calculations have shown that a limit 
at 300~Hz seems plausible in terms of neutron star physics.

This would make such systems steady gravitational wave beacons. The strongest 
would be the one whose X-ray flux on the Earth was strongest, since this correlates 
directly with the gravitational wave flux.  This system is Sco X-1, the first extra-solar-system 
X-ray source to have been discovered, and still one of the least understood.  

\begin{figure}
            \epsfxsize = 9 cm   
            \epsfclipon
            \centerline{\epsfbox{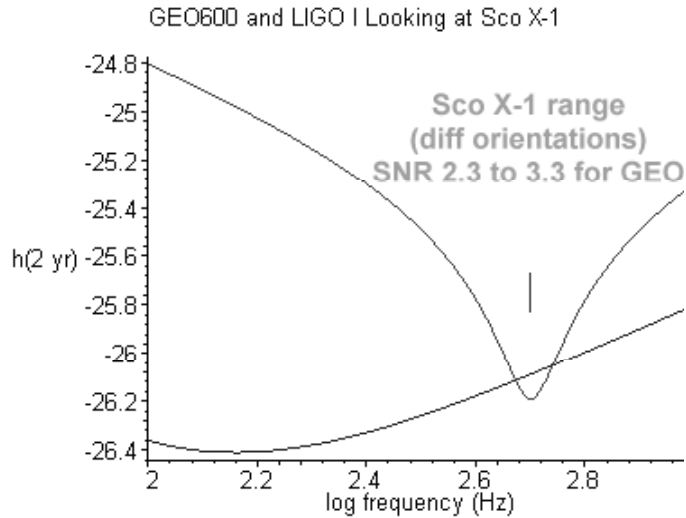}}
        \caption{Sensitivity of GEO600 and LIGO~I to the predicted 
radiation from Sco X-1, assuming perfect filtering during a two-year 
observation.  The antenna pattern effects have been taken into 
account.  The range of values for the signal reflect uncertainty about 
the inclination of the spin axis of Sco X-1 to the line of sight.}
        \label{fig:sco}
        \end{figure}

Figure~\ref{fig:sco} shows the sensitivity of GEO600 to the predicted radiation from Sco X-1 
if this radiation comes out at 500~Hz.  By narrow-banding, GEO600 can achieve slightly 
better sensitivity than LIGO~I, which will not be able to narrow-band.  Nevertheless, the 
optimum SNR is not high.  In a two-year observation with a perfectly matched filter, 
GEO600 will reach an SNR only of order 3.  In a single detector this would not be enough to 
identify the radiation.  But in three detectors (GEO600 plus the two LIGO detectors), 
the confidence could be sufficient.  

However, the construction of a good matched filter will not be easy. 
The accretion rate in such systems is variable, and this will lead 
to small variations in the spin rate and hence in the emitted 
frequency.  Taken over a long enough time, these may produce a 
random-walk-style drift in frequency that is observable.  It may be 
that monitoring of the emissions from Sco X-1 will provide enough 
information to make at least a sensibly parametrized model of 
the signal, but this remains to be seen.

For GEO600, the decision to look for Sco X-1 will be a difficult one, 
since it will involve dedicating the detector to a 2-year narrow-band 
observation.  This decision will have to be made in the light of 
further research on models for Sco X-1 and after it is clear whether 
a good filter can be developed.  

In the absence of such a good filter, testing the model for Sco X-1 in 
this way will have to wait for LIGO~II, which could detect the predicted 
radiation in an observation lasting a few days.

\subsection{ Bursts and unexpected signals}
Surprises, usually in the form of strong unexpected sources, 
have turned up in every new waveband that astronomy has 
looked in: radio galaxies, X-ray binaries, gamma bursts,
ultra-high-energy cosmic rays, infrared galaxies.  At some 
level of sensivity, this must also happen for gravitational 
waves.  We hope to increase the probability of finding 
unexpected sources early by adopting suitable signal-search 
methods.

It is of course not possible to perform matched filtering for 
a signal for which no waveform is predicted.  However, 
it is possible to argue plausibly that there are some features 
to be expected even from unexpected sources.  In 
particular, it seems to me that strong sources of 
radiation may be dominated by rotation, since rotation 
is a natural way to produce the kinds of asymmetry 
that are needed for gravitational radiation.  Strong 
sources should therefore consist of bursts of radiation 
that are at least a few cycles long, if not longer. This 
might be particularly true of bursts from supernovae.

There are signal-search methods that look for short-lived 
signals with relatively narrow bandwidth.  Linear methods 
such as wavelets and other time-frequency methods seem 
well-suited to this job.  Nonlinear adaptive filtering methods 
that lock onto oscillations may also be useful.  Both of these 
are under investigation within the GEO team and elsewhere, 
such as in the LSC.  These methods should also reveal 
short oscillatory features of instrumental noise, which may 
be abundant because of the number of feedback-control 
systems in the detectors.  They will therefore be used to 
clean up the noise or at least to veto sections of the data. 
The real problem will be to recognize which of these 
events, if any, comes from a gravitational wave instead 
of an internal system.  Coincidence analysis with 
data from other detectors using the same filters seems 
essential here.

\section{Conclusions }
The development of algorithms for the GEO600 data analysis 
is nearing completion in a number of areas, and software is 
being written with the goal of having a fully tested working system before 
the data rolls off in mid-2001.  The GEO600 analysis system 
will use components created in-house and components imported from 
the LIGO LSC.  Other collaborations would be most welcome. 
There will undoubtedly be gravitational wave signals in our 
first data, but will we be able to recognize them?  The detector 
builders are doing their best to ensure that there is as 
little noise obscuring them as possible using our current 
technology.  We data analysts must similarly ensure that we 
do the best possible job of filtering out the remaining noise 
numerically, as we look for recognizable signals.

\section*{Acknowledgements}
I would like to thank the GEO data analysis team
 and the GEO experimenters 
for many helpful conversations. I am also indebted to Bruce Allen, Patrick Brady, 
Teviet Creighton, and Kip Thorne for useful discussions on a number of points.

\end{document}